\documentstyle[12pt]{article}

\def\hybrid{\topmargin -20pt    \oddsidemargin 0pt
    \headheight 0pt \headsep 0pt
    \textwidth 6.25in       % A4 paper
    \textheight 9.5in       % A4 paper
    \marginparwidth .875in
    \parskip 5pt plus 1pt   \jot = 1.5ex}

\hybrid

\def\baselinestretch{1.4}
\begin{document}

%  Greek letters
\def\a{\alpha}
\def\b{\beta}
\def\ch{\chi}
\def\d{\delta}
\def\e{\epsilon}
\def\E{{\cal E}}
\def\f{\phi}
\def\Tr{\mbox{Tr}}
\def\g{\gamma}
\def\h{\eta}
\def\i{\iota}
\def\j{\psi}
\def\k{\kappa}
\def\l{\lambda}
\def\m{\mu}
\def\n{\nu}
\def\o{\omega}
\def\p{\pi}
\def\q{\theta}
\def\r{\rho}
\def\s{\sigma}
\def\t{\tau}
\def\u{\upsilon}
\def\x{\xi}
\def\z{\zeta}
\def\D{\Delta}
\def\F{\Phi}
\def\G{\Gamma}
\def\J{\Psi}
\def\L{\Lambda}
\def\O{\Omega}
\def\P{\Pi}
\def\S{\Sigma}
\def\U{\Upsilon}
\def\X{\Xi} 
\def\T{\Theta}
\def\Q{\Theta}
\def\vf{\varphi}
\def\ve{\varepsilon}
\def\cC{{\cal P}}
\def\cD{{\cal Q}}

\def\Ab{\bar{A}}
\def\gi{g^{-1}}
\def\li{{ 1 \over \l } }
\def\lb{\l^{*}}
\def\zb{\bar{z}}
\def\ub{u^{*}}
\def\vb{v^{*}}
\def\Tb{\bar{T}}
\def\pp {\partial }
\def\pb {\bar{\partial }}
\def\be{\begin{equation}}
\def\ee{\end{equation}}
\def\ben{\begin{eqnarray}}
\def\een{\end{eqnarray}}
\def\lt{\tilde{\lambda}}

\thispagestyle{empty}
\begin{flushright} March \ 1997\\
SNUTP 97-016 / hep-th/9703096 \\
\end{flushright}
\begin{center}
 {\large\bf Path Integral Bosonization of Massive GNO Fermions }
\vglue .5in
 Q-Han Park\footnote{ E-mail address: qpark@nms.kyunghee.ac.kr }
\vglue .2in
{and}
\vglue .2in
H. J. Shin\footnote{ E-mail address: hjshin@nms.kyunghee.ac.kr }
\vglue .2in
{\it  
Department of Physics \\
and \\
Research Institute of Basic Sciences \\
Kyunghee University\\
Seoul, 130-701, Korea}
\vglue .2in
{\bf ABSTRACT}\\[.2in]
\end{center}

We show the quantum equivalence between certain symmetric space sine-Gordon models 
and the massive free fermions. In the massless limit, these fermions reduce to 
the free fermions introduced by Goddard, Nahm and Olive (GNO) in association with 
symmetric spaces $K/G$. A path integral formulation is given in terms of the 
Wess-Zumino-Witten action where the field variable $g$ takes value in the orthogonal, 
unitary, and symplectic representations of the group $G$ in the basis of the 
symmetric space. We show that, for example, such a path integral bosonization is possible 
when the symmetric spaces $K/G$ are $ SU(N) \times SU(N)/SU(N); N \le 3,  ~ Sp(2)/U(2) $ 
or $ SO(8)/U(4)$. We also address the relation between massive GNO fermions and 
the nonabelian solitons, and explain the restriction imposed on the fermion mass 
matrix due to the integrability of the bosonic model.

\vglue .1in  
\def\baselinestretch{1.5}
\newpage 
\noindent               
{\bf 1. Introduction}
\vglue .1in
The quantum equivalence between the sine-Gordon and the massive Thirring 
models, which is known as abelian bosonization, is one of the noble features of 
two-dimensional quantum field theories \cite{coleman}. In the massless case, 
it was subsequently generalized by Witten \cite{witten} to the nonabelian bosonization, 
where the group  $SO(N)$ (or $U(N)$) Wess-Zumino-Witten model at 
level one is shown to be quantum equivalent to the free Majorana (or Dirac) 
fermion model. The fermion mass bilinears are identified with bosonic operators 
in the massless model; i.e., Witten's nonabelian bosonization considers 
only the zero charge sector of the fermionic model where massive excitations are 
treated perturbatively. This should be contrasted to the abelian case 
where the massive fermion operator (charge nonzero sector) is identified 
with a nonperturbative bosonic soliton operator \cite{mandelstam}. 
The nonabelian bosonization for the nonzero charge sector requires a 
generalization of the Mandelstam's soliton operator. 
Recently, a systematic nonabelian generalization of the sine-Gordon model has 
been made in association with symmetric spaces \cite{bakas1} and nonabelian 
soliton solutions are found which extend the sine-Gordon soliton with 
extra internal degrees of freedom \cite{shin2}. This raises a possibility of 
massive nonabelian bosonization for the nonzero charge sector in terms of 
nonabelian soliton operators. On the other hand, Goddard, Nahm and 
Olive (GNO) \cite{GNO} have shown that the free, massless fermions 
having Sugawara's energy-momentum tensor are in one to one correspondence 
with compact symmetric spaces. In the context of conformal embedding, this 
correspondence led to the quantum equivalence of the GNO's free fermions 
to the Wess-Zumino-Witten (WZW) models. However, for the quantum equivalence, 
it was pointed out that having the same Sugawara's energy-momentum tensor 
(or Virasoro symmetry) alone was not sufficent, but it also requires the same spectra 
of primary fields and the same operator product algebras \cite{fuchs}. 
In general, the equivalent bosonic model is not described by a (diagonal) WZW model 
on a simply connected group manifold, and sor far, a path integral formulation of such 
bosonic model has been lacking. 
Only the trivial conformal embeddings, $\widehat{so(N)}_{k=1} \subset  
\widehat{so(N)}_{k=1}$ and  $\widehat{su(N)}_{k=1} \subset  \widehat{su(N)}_{k=1}$,
admit a path integral bosonization of free Majorana and Dirac fermion theories 
in terms of diagonal WZW actions which is precisely the Witten's bosonization \cite{witten}.
This makes particularly difficult the association of GNO fermions with bosonic solitons.

The purpose of this letter is to show that, with certain restrictions, 
symmetric space sine-Gordon models indeed bosonize the massive nonabelian fermions. 
In the massless limit, these fermions become GNO fermions.  
In particular, we present a path integral bosonization of massive GNO fermions
in terms of a WZW action, where the field variable $g$ takes value in the 
orthogonal, unitary, and symplectic representations of the group $G$ in the basis 
of the tangent space generators of the symmetric space $K/G$. We construct 
representations explicitly, and specify the kernels of each representation
for various symmetric spaces. 
We show that certain GNO fermions admit a path integral bosonization when the partition 
function belongs to the A or D-series of modular invariants \cite{cappeli}. 
For example, if we consider the type II symmetric space; $K/G = SU(N) \times SU(N)/SU(N)$, 
it corresponds to the bosonization of $SO(3)$ Majorana fermions for $N=2$ and  of 
$SO(8)$ Majorana fermions for $N=3$.  
Similarly, we show that the type I symmetric spaces, $ Sp(2)/U(2)$ and $ 
 SO(8)/U(4)$, also admit a path integral bosonization. 
Identifying the bosonic model with integrable symmetric space sine-Gordon models,     
we explain the relation between the massive GNO fermions and 
the nonabelian solitons, and explain the restriction imposed on the fermion mass 
matrix due to the integrability of the bosonic model.
 \vglue .1in  
\noindent
{\bf 2. Orthogonal representations}
\vglue .1in
Consider the symmetric space $K/G$ for Lie groups $G \subset K$ whose associated Lie 
algebras ${\bf g} \subset {\bf k}$  satisfy the Lie algebra commutation 
relations;                                       
\be
[ \bf{g} \ , \ \bf{g} ] \subset \bf{g} \ ,\ [ \bf{g} \ , \ \bf{p} ] 
\subset \bf{p} \ , \ 
[\bf{p} \ , \ \bf{p} ] \subset \bf{g} \ .
\ee
Here, $\bf{p}$ is the vector space complement of $\bf{g}$ in $\bf{k}$, i.e., 
\be
\bf{k} = \bf{g} \oplus \bf{p}.
\label{decom}
\ee   
We denote orthogonal basis of $\bf{g} $ and $\bf{p}$ by $T_i $ and $P_\a $ and 
assume that
\be
[T_i ~, ~ T_j ] = if_{ijk}T_{k}~~, ~~ [T_i ~, ~ P_\a ] = iP_\b M^{i}_{\b \a }   .
\label{struct}
\ee
We scale $T_{i}$ and $P_\a $ such that
\be
\Tr (T_i T_j) = y \d_{ij},~~  \Tr (P_\a P_\b) = y \d_{\a \b} .
\ee
Then, GNO's theorem states that free fermions possess Sugawara's energy-momentum 
tensor if free fermions transform under $G$ as the generators $P_\a $ of the 
symmetric space $K/G$ do. These free fermions will be called as GNO fermions. 
The Sugawara's energy-momentum tensor of GNO fermions also realize the conformal 
embedding ${\bf \hat{g}} \subset {\bf \hat{f}}$ of affine Lie algebras, that is,
${\bf \hat{g}} $ and $ {\bf \hat{f}}$ resulting in the same Virasoro algebra. 
Here,  ${\bf \hat{f}}$ is the Kac-Moody algebra associated with the embedding 
group $F = SO(\mbox{dim } {\bf p} )$ (or $U(\mbox{dim } {\bf p}/2 ), 
Sp(\mbox{dim } {\bf p}/4 ) )$ as will be explained later.
In fact, it was shown that all possible subalgebras ${\bf \hat{g}} $ of affine 
algebras ${\bf \hat{f}} = \widehat{so(n)}, \widehat{su(n)}$ or $\widehat{sp(n)} $, 
satisfying the conformal embedding condition, can be found directly from the 
known classification of symmetric spaces \cite{arcuri}.

In order to make the conformal embedding explicit, we represent the embedding 
$G \subset F$ using the orthogonal representation of $g \in G$ 
in the basis of the symmetric space generators $P_\a $ as follows;                                                                                    
\be
f_{\a \b} \equiv {1 \over y} \Tr g^{-1} P_\a g P_\b.
\label{ortho}
\ee
For the case where $g^{-1} = g^\dagger$ and 
$P_\a ^\dagger = -P_\a$, it is easy to see that matrices 
$f_{\a \b}$ are real and becomes an element of a subgroup of 
$F=SO(\mbox{dim \bf p})$. 
Thus, it provides an orthogonal matrix representation of the group $G$ 
embedded in $F$ which, in general, is reducible. 
For the Hermitian symmetric spaces ({\bf AIII, BDI, CI, DIII, EIII, EVII} type),
the embedding can be restricted to a unitary representation as follows;
a Hermitian symmetric space, equipped with a complex structure, 
has a commuting $U(1)$ factor $e^{\q T}$ in $G$, i.e., $G=e^{\q T} \times G^{'}$ and 
$ [T, ~ G] = 0$. $T$ is an element of the Cartan subalgebra of 
${\bf k}$ whose adjoint action introduces a $Z_{2}$-grading over ${\bf p}$. 
Thus, under the adjoint action of $T$, generators of ${\bf p}$ can be regrouped
into a set of pairs $(P_\a ^{(1)}, P_\a^{(2)}); ~ 
\a=1, \cdots , \mbox{dim } {\bf p}/2,$ such that
\be
[T,~ P_\a^{(1)}]=P_\a ^{(2)}, ~~[T,~ P_\a ^{(2)}]=-P_\a ^{(1)}.
\ee                                                                     
In this new basis, the $ U(1)$ factor $e^{\q T}$ can be represented by a block 
diagonal matrix where each block is given by the $2 \times 2$ matrix, 
$\pmatrix{ \cos \q & -\sin \q \cr \sin \q & \cos \q } $. 
Since $ G$ commutes with the $U(1)$ factor, a group element of $G$ takes 
a form of the ${ 1\over 2}\mbox{dim } {\bf p} \times { 1\over 2}\mbox{dim } {\bf p}$ 
matrix whose entries are given by $2 \times 2 $ matrices of the form 
$\pmatrix{ x & -y \cr y & x }$. Or, equivalently, by 
a complex number $x + i y$. Therefore, in the case of Hermitian symmetric spaces, the 
orthogonal representation reduces to the unitary representation with
$F= U(\mbox{dim } {\bf p}/2 )$, similar to the complex (Dirac) representation 
of fermions instead of the real (Majorana) representation.
As an example, consider the case $K/G =  SU(3) / ( SU(2) \times U(1))$. 
Among eight Gell-Mann matrices $\l_i$, the element $T$ is given by $\l_8 /
\sqrt{3} i $ and $\l_1, \l_2 $ and $ \l_3$ form an $su(2)$ subalgebra. 
Also, we have $P_1^{(1)} = \l_4,~  P_1^{(2)} = \l_5, ~ P_2^{(1)} = 
\l_6, ~ P_2^{(2)} = \l_7$ so that ${\bf g}=\{ \l_{1} , \l_{2}, \l_{3}, T \} $ 
and $ {\bf p} = \{ \l_{4} , \l_{5}, \l_{6}, \l_{7} \} $. 
In this way, $G=SU(2) \times U(1)$ becomes conformally 
embedded into the unitary group $F=U(2)$.

Similarly, for the symmetric spaces ({\bf AIII, BDI}), the orthogonal representation 
reduces to the simplectic one, with $F=Sp(\mbox{dim } {\bf p}/4 )$. 
In this case, group $G$ contains $SU(2)$  or $SO(3)$ factors which commutes with $G$.
We may again choose a basis of the symmetric space so that generators $P_\a $ 
can be regrouped into a set of quadruples $(P_\a ^{(1)}, P_\a ^{(i)}, P_\a ^{(j)},  
P_\a ^{(k)}); \a=1, \cdots , \mbox{dim } {\bf p}/4 ,  $ labeled by quaternion numbers. 
They satisfy the commutation relations; 
\be
[T_A, P_\a ^{(B)}] = \e_{ABC} P_\a ^{(C)} ~ \mbox{ for } A = i, j, k , ~~ \mbox{ and }
 B = 1, i, j , k
\label{comm}
\ee
where $T_{i}, T_{j}$ and $ T_{k} $ are the generators of $su(2)$ or $so(3)$.
The symbol $\e$ stands for  
\be
\e_{ABC} = \d_{AB , C}
\ee
where $AB$ in $\d_{AB , C}$ denotes the quaternion product and the dummy index 
$C$ runs over $\pm 1 , \pm i, \pm j , \pm k$. 
For instance,  $\e_{i1i}= \e_{ijk} = 1, ~
\e_{kji} =  \e_{jj1} = -1, ~ \e_{iij}= \e_{ij1}= 0 $ etc..
In this case, the $SU(2)$ or $SO(3)$ factor can be represented by a block diagonal 
matrix where each block is equal to  the $4 \times 4$ matrix
$\sum_{i=0}^{3} \r_i L_i $ with real $\r_{i}$ satisfying $\sum \r_i ^2 = 1$.
The matrices $L_i $ are
\be
L_0 = \left( \begin{array}{cc} 1 & 0 \\ 0 & 1 \end{array} \right),
L_1 = \left( \begin{array}{cc} i \s_2 & 0 \\ 0 & i \s_2 \end{array} \right),
L_2 = \left( \begin{array}{cc} 0 & \s_3 \\ -\s_3 & 0 \end{array} \right),
L_3 = \left( \begin{array}{cc} 0 & \s_1 \\ -\s_1 & 0 \end{array} \right)
\ee
where $\s_i$ are Pauli matrices.
Since $G$ commutes with the $SU(2)$ or $SO(3)$ factor, $G$ takes a form of 
the $(\mbox{dim } {\bf p}/4) \times (\mbox{dim } {\bf p}/4) $ matrix whose entries
are $4 \times 4 $ matrices commuting with $L_i$. Thus, it takes a form,
$\sum_{i=0}^3 a_i M_i$, for real $a_i $ where
\be
M_0 = \left( \begin{array}{cc} 1 & 0 \\ 0 & 1 \end{array} \right),
M_1 = \left( \begin{array}{cc} i \s_2 & 0 \\ 0 & - i \s_2 \end{array} \right),
M_2 = \left( \begin{array}{cc} 0 & 1 \\ -1 & 0 \end{array} \right),
M_3 = \left( \begin{array}{cc} 0 & i\s_2 \\ i\s_2 & 0 \end{array} \right) .
\ee
Or, equivalently, it may be identified with a quaternion number $a_0 + a_1 i
+ a_2 j + a_3 k$ so that the orthogonal representation reduces to the 
simplectic matrix representation.
In the GNO's fermionic theory, this case corresponds to taking the complex 
representation of quark multiplets pseudoreal, i.e. it is equivalent to its 
complex conjugate but not real.
As an example, we consider the $SU(N+2) /( SU(2) \times U(N))$ case. 
Then, we have
\be
T_{i} = \pmatrix{0 & -i &  \cdots & 0 \cr 
                 -i & 0 &           &   \cr
                  \vdots &&&          \vdots \cr 
                  0 & \cdots && 0 }, ~~         
T_{j} = \pmatrix{0 & -1 &  \cdots & 0 \cr 
                 1 & 0 &           &   \cr
                  \vdots &&&          \vdots \cr 
                  0 & \cdots && 0 }, ~~         
T_{k} = \pmatrix{-i & 0 &  \cdots & 0 \cr 
                 0 & i &           &   \cr
                  \vdots &&&          \vdots \cr 
                  0 & \cdots && 0 }         
\ee
and
\ben  
P_{\a - 2}^{1} &=&  
\bordermatrix{
               &  &\cdots& \a  &\cdots\cr
               &  &      & -i&    \cr 
         \vdots&  &      & 0 &   \cr
            \a &-i& 0     & \cdot & \cdots     \cr
         \vdots& &     &\vdots  &      \cr}  , ~~
P_{\a - 2}^{i} =  
\bordermatrix{
               &  &\cdots& \a  &\cdots\cr
               &  &      & 0&    \cr 
         \vdots&  &      &-1&   \cr
            \a &0&1      & \cdot & \cdots     \cr
         \vdots& &     &\vdots  &      \cr} 
 ~; ~~ 3\le \a \le N+2       \nonumber \\
P_{\a - 2}^{j} &=&  
\bordermatrix{
               &  &\cdots& \a  &\cdots\cr
               &  &      & 0&    \cr 
         \vdots&  &      &-i&   \cr
            \a &0& -i      & \cdot & \cdots     \cr
         \vdots& &     &\vdots  &      \cr} , ~~
P_{\a - 2}^{k} =  
\bordermatrix{
               &  &\cdots& \a  &\cdots\cr
               &  &      & -1&    \cr 
         \vdots&  &      & 0&   \cr
            \a &1& 0      & \cdot & \cdots     \cr
         \vdots& &     &\vdots  &      \cr}               .
\een
These $T_A$ and $P_{\a }^{B}$ satisfy the commutation  relation Eq. (\ref{comm}) 
and the representation becomes a simplectic one with $F= Sp(N)$. 
\vglue .1in                                 
\noindent
{\bf 3. Path integral bosonization}
\vglue .1in
Having specified the orthogonal representation of the group $G$ as in Eq. (\ref{ortho}), 
we now consider the level one WZW model given by the path integral, 
\be
\int [df]\exp[S_{WZW}(f)]      ,
\label{model}
\ee
where the field $f$ is given by Eq. (\ref{ortho}) and $S_{WZW}$ is the usual WZW action.
In general, the orthogonal representation is reducible. Notable exceptions are  
when $K/G  = SO(N+1)/SO(N) = S^N $ or $K/G = SU(N+1)/U(N) 
= CP^N $. In such cases, the orthogonal representation becomes the defining 
fundamental representations of $SO(N)$ and $U(N)$ respectively. 
Also, the model in Eq. (\ref{model}) expresses the trivial conformal embeddings, 
${\bf \hat g} = \widehat{so(N)}_{k=1} \subset {\bf \hat f} =  
\widehat{so(N)}_{k=1}$ and  $\widehat{su(N)}_{k=1} \subset  
\widehat{su(N)}_{k=1}$, which are equivalent respectively to the Majorana and Dirac 
fermionic models \cite{witten}. However, for other cases of symmetric spaces, 
the embedding becomes nontrivial. The currents of the orthogonal representation, 
e.g. the chiral Kac-Moody current, are related to those of the defining representation 
in the following way;
\ben
(f^{-1} \pp f)_{\a\b} &=& {1 \over y} \Tr(g P_\a g^{-1} P_\g )
\Tr ([g^{-1} P_\g g, g^{-1} \pp g] P_\b ) \nonumber \\ 
&=& {1 \over y} \Tr ( P_\a [g^{-1} \pp g, P_\b] )  \nonumber \\
&=& i M^{i}_{\a\b} (g^{-1} \pp g)_i .
\label{vari}
\een
The structure constants $M^{i}_{\a\b}$ are as given in Eq. (\ref{struct}) and 
possess the property,
\be
\sum_{\a\b} M^i_{\a\b} M^j_{\a\b} = \k \d_{ij} , 
\ee                                              
where  
$\k= x \j^2, \j = \mbox{highest root of } g$ and $x$ is the Dynkin index 
of the representation with generators $M^i$. 
Thus, the bosonization of the fermionic current $J^i ={i \over 2} 
\j^{\a} M^{i}_{\a\b} \j^{\b}$ is that
\be
J^i = M^{i}_{\a\b} (f^{-1} \pp f)_{\a\b} = i \k (g^{-1} \pp g)_i   ,
\label{current}
\ee
and a straightforward calculation shows that $J^i$ satisfies the 
Kac-Moody algebra of group $G$ with the level $\k/2$.
Since Eq. (\ref{vari}) holds also for the case where $f^{-1}\pp f $ and $g^{-1} \pp g$ 
are replaces by any infinitesimal variations $f^{-1}\d f $ and $g^{-1} \d g$, 
the WZW action reduces to 
\be
S_{WZW}(f) = {\k \over 2} S_{WZW}(g) . 
\label{red}
\ee
However, it should be emphasized that the identity, Eq. (\ref{red}), implies 
the equivalence of the model in Eq. (\ref{model}) with the level $\k $, group $G$ 
WZW model only at the classical level. At the quantum level, 
since the orthogonal, as well as the unitary and the symplectic, representation 
of group $G$ is not necessarily faithful, the model in Eq. (\ref{model}) is not in 
general equivalent to the level $\k $, diagonal WZW model based on the simply 
connected group $G$.  
For example, the orthogonal represenation of a type II symmetric space, 
$G \times G/G,$  is the same as the adjoint representation of $G$, and the kernel of 
the adjoint representation is the center $Z_{G}$ of the group $G$. 
Thus, the model in Eq. (\ref{model}) corresponds to the group $G$ WZW model modded out by 
the center $Z_G$. More specifically, if $G = SU(2)$, the model in Eq. (\ref{model}) 
possesses the level two Kac-Moody algebra $\widehat{su(2)}_{k=2}$  and the 
orthogonal representation corresponds to $SO(3) \approx SU(2)/Z_{2}$. Even though, 
Kac-Moody algebras corresponding to $SU(2)$ and $SO(3)$ are equivalent, the 
corresponding WZW partition functions are different \cite{gepner}.  Thus, the model 
in Eq. (\ref{model}) truely becomes equivalent to the level one $SO(3)$ WZW model, 
or through the Witten's bosonization, to the $SO(3)$ massless Majorana fermions. 
This provides a field theoretic realization of the conformal embedding, 
$\widehat{su(2)}_{k = 2} \subset \widehat{so(3)}_{k=1} $.  
Previously, such Majorana fermions have also been used in representing the Kac-Moody 
algebra, $\widehat{su(2)}_{k=2} $\cite{zam}.    
Similarly, the model Eq. (\ref{model}) for the type II space with $G = SU(3)$ 
realizes the conformal embedding $\widehat{su(3)}_{k=3} \subset 
 \widehat{so(8)}_{k=1} $. That is, the partition of the model in Eq. (\ref{model}) in this 
case is that of the level three $SU(3)$ WZW model modded by $Z_{3}$, 
which is $Z(A_{6}/Z_{3})$ in the $D$ series of modular invariants \cite{cappeli}
(see e.g. Ref. \cite{abol} for the notation).
On the other hand, this is also the partition function of the level one $SO(8)$ WZW 
model thus proving the equivalence of the model in Eq. (\ref{model}) with the $SO(8)$ Majorana 
fermions. However, for $G = SU(N), N \ge 4 $, the relevant conformal embeddings belong 
to the $E$-series. For example, if $N=4$, the relevant 
conformal embedding, $\widehat{su(4)}_{k=4} \subset \widehat{(B_{7})}_{k=1}$,  
belongs to the exceptional case $E_{8}$. 
We do not know whether the model in Eq. (\ref{model}) still realizes the conformal 
embedding for these exceptional cases. This remains as an open problem.

As for the type I symmetric spaces, kernels of orthogonal representations can be 
computed through an explicit parametrization of each symmetric spaces. The results 
are the following; for symmetric spaces  {\bf AI} $ (F/G = SU(N)/SO(N)) $, ~ {\bf CI} 
$(Sp(N)/U(N))$ and  {\bf DIII} $(SO(2N)/U(N))$, kernels are $Z_{2}$ for even $N$. 
For odd $N$, they become trivial.
For symmetric spaces {\bf AII} $(SU(2N)/Sp(N))$ and {\bf CII} $(Sp(M+N)/(Sp(M) \times Sp(N)))$, 
kernels are $Z_{2}$ while the kernel of {\bf AIII} 
$(SU(M+N)/(SU(M) \times SU(N) \times U(1)))$ case is  
$Z_{(M,N)}$ where $(M,N)$ denotes the greatest common divisor of $M$ and $N$. 
For {\bf BDI} $(SO(M+N) /(SO(M) \times SO(N))) $, it becomes $Z_{2}$ for $M$ and $N$ all 
even and otherwise it becomes trivial. We have not computed kernels for the exceptional 
cases leaving it for a future work. With kernels specified, we could check the 
quantum equivalence of the model in Eq. (\ref{model}) with free fermion models. 
For instance, the $ N=2$ case of the {\bf CI} symmetric space corresponds to the conformal 
embedding $\widehat{su(2)}_{k=4 } \subset \widehat{su(3)}_{k=1}$. The partition function 
is $Z(A_{6}/Z_{2})$ in the $D$-series which shows the equivalence of the model in Eq. (\ref{model}) with 
$U(3)$ Dirac fermions. However, in the $N=3$ and $N=4$ cases, 
the relevant embeddings are $\widehat{su(3)}_{k=5 } \subset \widehat{su(6)}_{k=1}$ and 
$\widehat{su(4)}_{k=6 } \subset \widehat{su(10)}_{k=1}$ respectively, and partition 
functions belong to $E_{8}$ and $E_{10}$. Another example is the $N=4$ case of {\bf DIII} 
symmetric space which has the conformal embedding $\widehat{su(4)}_{k=2 } \subset 
\widehat{su(6)}_{k=1}$ with the partition function $Z(A_{6}/Z_{2}) = D_{6}$ thereby 
revealing the quantum equivalence of the model in Eq. (\ref{model}) with $U(6)$ Dirac fermions.
The $N \ge 5$ cases of the {\bf DIII} symmetric space are again associated with  
the $E$-series so that the quantum equivalence with the fermion model is not clear. 
\vglue .1in
\noindent
{\bf 4. Fermion mass and nonabelian soliton} 
\vglue .1in 
Earlier attempts to extend the massive abelian bosonization to the nonabelian case 
resorted to $N$ scalar fields \cite{hal,banks}. However, in this approach, the 
global nonabelian symmetry of fermions become obscure and the bosonic off-diagonal 
currents become nonlocal. In the Witten's bosonization, such difficulties were removed by 
introducing a nonlinear sigma field $g$ in terms of which bosonization is done in 
a local way with manifest global symmetries. On the other hand, the existence of 
solitons in the bosonized model, which generalizes the sine-Gordon soliton, has not 
been well understood. In the spirit of particle vs. soliton duality in the 
abelian bosonization, the bosonic solitons associted with nonabelian fermions can be 
of interest. In the following, we show how such nonabelian solitons can 
arise in the model which bosonizes massive GNO fermions. 

In the previous section, we have shown that the model in Eq. (\ref{model}) for 
specific symmetric spaces become equivalent to Majorana or Dirac fermions 
according to the orthogonal or unitary representations of embedding. 
We note that the bosonic bilinear in $g$ can be identified with 
a fermion bilinear in our case;
\be
{1 \over y}\mbox{Tr} g^{-1}P_{\a}gP_{\b}
 ~ \equiv f_{\a\b }(z, \bar{z}) = M \j^{\a }(z) \bar{\j }^{\b }(\bar{z})
\label{mass}
\ee
where $M$ is the mass parameter dependent on the regularization scheme.
Thus, the bosonic model in Eq. (\ref{model}) with a bosonized fermion 
mass term is given by
\ben
S &=& S_{WZW}(f) + \int M_{\a\b }f^{\a\b} \nonumber \\
 &=& {\k \over 2}S_{WZW}(g) + \int M_{\a\b }\mbox{Tr}({1 \over y} 
  g^{-1} P^{\a} g P^{\b} ). 
\label{model2}
\een
In general, this model is not integrable. However, when the mass matrix 
$M_{\a\b}$ takes a specific form, e.g. if only one component is nonvanishing, e.g. 
$M_{AB} \ne 0$, then the classical equation of motion of Eq. (\ref{model2}) 
takes a zero curvature form, 
\be
[\pp + g^{-1}\pp g + m\l P^{B} ~ , ~~ \pb + {1\over \l }g^{-1}P^{A }g ] =0 ,
\ee
where $\l $ is a spectral parameter and $m$ is a mass parameter. This 
demonstrates the integrability of the model, at least classically \cite{park}. 
With a particular choice of $ P^{A}$ and $ ~ P^{B }$,  also with additional constraints 
which removes the massless degrees of freedom, this model has been called as a 
symmetric space sine-Gordon model \cite{regge}-\cite{hollo} and its various properties 
including nonabelian solitons have been investigated \cite{shin}-\cite{shin2}.

In order to understand the existence of solitons, we assume that $M_{ 11 } \ne 0 $ and all 
other components of $M_{\a\b}$ are zero. The 
potential term $V = M_{11}f^{11} = M_{11}\mbox{Tr}(g^{-1} P^{1} g P^{1} ) /y $ is 
in general quadratic in $g$. However, in the case where $K/G  = SO(N+1)/SO(N) = S^N $ 
or $K/G = SU(N+1)/U(N) = CP^N $, the potential $V$ becomes $V = \mbox{const.} (g_{11} + 
g^{-1}_{11})$. Since $g$ is unitary, it satisfies
\be 
|g_{11}|^2 +|g_{12}|^2 + \cdots +
 |g_{1N}|^2 =1 .
\ee
Thus, we could write $g_{11}$ by $g_{11} = \cos {\phi }e^{i \q}$ for some scalar fields 
$\phi $ and $\q $ so that the potential $V= \mbox{const.} \cos{\phi}\cos{\q}$. This shows 
that the bosonized fermion mass term induces infinitely many degenerate vacua, and 
essentially soliton solutions interpolating different vacua.  For concreteness, we write 
the Lagrangian of the massive bosonic model in Eq. (\ref{model2}) for the $CP^2$ case 
in terms of three scalar fields $\phi , \q_{1} $ and $\q_{2}$,
\be
L = {1 \over 2\pi }(\pp \phi \pb \phi + {1 \over 4 }\pp \q_{1}\pb \q_{1} + \cot^{2}{\phi }\pp \q_{2} \pb \q_{2} 
+ 2m \cos{\phi }\cos(\q_{1} - \q_{2}))           .                                        
\label{lagr}
\ee  
A consistent reduction can be made by taking 
$\q_{1} = \q_{2}$ and then the model reduces the complex sine-Gordon model, which is known to possess 
both topological and non-topological solitons \cite{lund,pohl}. 
Of course, a further consistent reduction 
which takes $\q_{1}=\q_{2}=0$ brings the Lagrangian in Eq. (\ref{lagr}) to that of the sine-
Gordon model. One can readily check that in this case the coupling constant of the sine-Gordon 
model is fixed in such a way that the corresponding massive fermion does not have Thirring 
interaction term \cite{coleman}. This shows that the fermion operator $\psi_{1}$ can be expressed in terms of the 
Mandelstam's soliton operator. The existence of non-topological solitons in the complex 
sine-Gordon model and also other localized solutions with  nonabelian nature suggest that 
the particle v.s. soliton duality could be further carried out in a nonabelian context. This 
will be considered elsewhere.

\vglue .3in 
{\bf ACKNOWLEDGEMENT}
\vglue .2in
This work was supported in part by the program of Basic Science Research, 
Ministry of Education BSRI-96-2442, and by Korea Science and Engineering 
Foundation through CTP/SNU.
\vglue .2in

\end{document}